\documentclass[twocolumn,showpacs,preprintnumbers,amsmath,amssymb]{revtex4}
\usepackage{epsfig}
\usepackage{graphicx}
\usepackage{dcolumn}
\usepackage{bm}

\def\lsim{\buildrel < \over {_{\sim}}}
\def\gsim{\buildrel > \over {_{\sim}}}

\newcommand{\beq}{\begin{equation}}
\newcommand{\eeq}{\end{equation}}
\newcommand{\be}{\begin{eqnarray}}
\newcommand{\ee}{\end{eqnarray}}
\newcommand{\qt}{{\widetilde q}}
\newcommand{\Lmunu}{L_{\mu\nu}}
\newcommand{\WmunuA}{W^{\mu\nu}}

\begin{document}
\title{Electron- and neutrino-nucleus scattering in the impulse approximation 
regime}
\author{Omar Benhar$^{1,2}$}
\author{Nicola Farina$^{2}$}
\author{Hiroki Nakamura$^{3}$}
\author{Makoto Sakuda$^{4}$}
\author{Ryoichi Seki$^{5,6}$}%
\affiliation
{
$^1$ INFN, Sezione di Roma. I-00185 Roma, Italy\\
$^2$ Dipartimento di Fisica, Universit\`a ``La Sapienza''. I-00185 Roma, Italy \\
$^3$ Department of Physics, Waseda University, Tokyo 160-8555, Japan\\
$^4$ Department of Physics, Okayama University, Okayama, 700-8530, Japan\\
$^5$ Department of Physics, California State University, Northridge, 
California 91330, USA\\
$^6$ W.K. Kellog Radiation Laboratory\\
California Institute of Technology, Pasadena, California 91125, USA\\
}
\date{\today}
\begin{abstract}
A quantitative understanding of the weak nuclear response is a prerequisite 
for the analyses of neutrino experiments such as K2K and MiniBOONE, 
which measure energy and angle of the muons produced in neutrino-nucleus interactions
 in the energy range $0.5-3$ GeV and reconstruct the incident neutrino energy to 
determine neutrino oscillations. In this paper we discuss theoretical calculations 
of electron- and neutrino-nucleus scattering, carried out within the impulse 
approximation scheme using realistic nuclear spectral functions.
Comparison between electron scattering data and the calculated inclusive cross 
section off oxygen, at beam energies ranging between 700 and 1200 MeV, show that 
the Fermi gas model, widely used in the analysis of neutrino oscillation experiments, 
fails to provide a satisfactory description of the measured cross sections, and 
inclusion of nuclear dynamics is needed.
\end{abstract}
\maketitle

\section {Introduction}

The field of neutrino physics is rapidly developing after
atmospheric neutrino oscillations and solar neutrino oscillations
have been established \cite{Kajita,Solar,KamLAND}. Recently, the SK
Collaboration has found evidence of the oscillatory signature
in atmospheric neutrinos, improving the determination of $\Delta
m^2$ \cite{SKATM}, and K2K experiment has confirmed the 
oscillations of atmospheric neutrinos at 99.995\% CL
\cite{K2K,Nakaya}. These neutrino experiments measure energy
and angle of muons produced in neutrino-nucleus interactions and
reconstruct the incident neutrino energy, which determines the neutrino
oscillations. K2K took data in the $E_{\nu} =0.5-3$ GeV region, 
and the recent L/E analysis of the SK atmospheric neutrinos is
mainly based on the dataset extending from 0.5 to 25 GeV.
JPARC and NuMI neutrino
experiments \cite{JPARC,Nova} propose to measure
$\nu_{\mu} \rightarrow \nu_e $ oscillations and determine
$\Delta m^2$ with $1\%$ accuracy and $\sin^2 2\theta
_{13}$ above 0.006, using a narrow-band neutrino beam at
$E_{\nu}=0.8$ GeV (JPARC) and 2.0 GeV (NuMI, off-axis).

In view of these developments, 
it is vital that theoretical calculations of cross
sections and spectra achieve an accuracy comparable to the experimental
one, which in turn requires that the nuclear response to 
weak interactions be under control at a quantitative level.

 At $E_{\nu}$=3 GeV or less, quasi-elastic scattering and quasi-free
$\Delta$ production are the dominant neutrino-nucleus processes. 
However, reactions in this energy regime are associated with a 
wide range of momentum transfer, thus involving different aspects of
nuclear structure. 

Four decades of electron-nucleus scattering experiments have
unequivocally shown that the mean-field approximation, 
underlying the nuclear shell model, does not provide a fully
quantitative account of the data (see, e.g., Ref. \cite{omar_nuint04}
and references therein). When the momentum transfer involved is large, 
dynamical nucleon-nucleon (NN) correlations are known to be important,
 and a description of nuclear structure beyond the mean-field picture is needed. 
On the other hand, neutrino-nucleus reactions also occur, in fact rather
appreciably, with a small momentum transfer. Comparison between 
the data at $Q^2 <0.2\ {\rm GeV}^2$ and the predictions of the 
Fermi gas (FG) model \cite{Smith}, showing a sizable deficit of 
events \cite{Nakaya,Ishida}, 
suggests that a more realistic description of both nuclear properties and
the reaction mechanism is indeed required.

In this paper we discuss the extension of the many-body theory
of electron-nucleus scattering (see, e.g., Ref. \cite{vijay_nuint01}
and references therein) to the case of neutrino-induced reactions. 
We focus on the energy range $0.7-1.2$ GeV and analyze inclusive 
scattering of both electrons and neutrinos off oxygen,
the main target nucleus in SK, K2K and other experiments. 
The quasi-elastic and quasi-free $\Delta$ production cross sections
obtained from the FG model \cite{Smith,Seki} 
are compared to the results of the many-body approach developed in 
Ref. \cite{gofsix}, extensively used to analyze electron 
scattering data at beam energy up to few GeV \cite{gofsix,bp,bffs}.
Preliminary versions of the materials in this paper have appeared in 
the Proceedings of NuInt04 \cite{omar_nuint04,seki_nuint04,farina_nuint04}.

Section \ref{eA-formalism} is devoted to a summary of the formalism employed to 
calculate the electron-nucleus cross section at high momentum transfer, as well 
as to the discussion of the main ingredients entering its 
definition: the nuclear spectral function, the elementary 
cross section describing electron scattering off a {\it bound} 
nucleon and the folding function embodying the main effects of
final state interactions.

In Section \ref{eA-results} the results of our approach are compared to inclusive 
electron scattering data at $0.2 \lsim Q^2 \lsim 0.6$ GeV$^2$, while 
in Section \ref{nuA} we outline the extension of the formalism to the case
of charged current neutrino-nucleus scettering.
Finally, our conclusions are stated in Section V.

\section{Many-body theory of the electroweak nuclear response}
\label{eA-formalism}

\subsection{Electron-nucleus cross section}

The differential cross section of the process
\beq
e + A \rightarrow e^\prime + X \ , 
\label{process:e}
\eeq
in which an electron carrying initial four-momentum $k\equiv(E_e,{\bf k})$
scatters off a nuclear target to a state of four-momentum
$k^\prime\equiv(E_e^\prime,{\bf k}^\prime)$, the target final state 
being undetected, can be written in Born approximation as (see, e.g., 
Ref. \cite{IZ})
\beq
\frac{d^2\sigma}{d\Omega_{e^\prime} dE_e^\prime}=\frac{\alpha^2}{Q^4} 
\frac{E_e^\prime}{E_e}  \ \Lmunu\WmunuA \ ,
\label{e:cross:section}
\eeq
where $\alpha$ is the fine structure constant and 
$Q^2 = -q^2 = {\bf q}^2 - \nu^2$, $q = k - k^\prime \equiv(\nu,{\bf q})$
being the four momentum transfer. 

The leptonic tensor, that can be written, neglecting the lepton mass, as
\beq
\Lmunu = 2 \left[ k_\mu k_\nu^\prime + k_\nu k_\mu^\prime -g_{\mu\nu} (k k^\prime) \right] \ ,
\eeq
is completely determined by electron
kinematics, whereas the nuclear tensor $\WmunuA$ contains all the information 
on target structure. Its definition involves the initial
and final hadronic states $ |0\rangle$ and $|X \rangle$, carrying 
four-momenta $p_0$ and $p_X$, respectively, as well as the nuclear 
electromagnetic current operator
$J^\mu$:
\beq
\WmunuA = \sum_X \langle 0 | J^\mu | X\rangle 
 \langle X | J^\nu | 0\rangle \delta^{(4)}(p_0+q-p_X) \ ,
\label{e:hadrten}
\eeq
where the sum includes all hadronic final states.

Calculations of $\WmunuA$ at moderate momentum transfers 
$( {\bf |q|} < 0.5\, {\rm GeV})$
can be carried out within nuclear many-body theory (NMBT), using
nonrelativistic wave functions to describe the initial and final
states and expanding the current operator in powers of ${\bf |q|}/m$, 
$m$ being the nucleon mass (see, e.g., Ref. \cite{rocco}). On the other hand, 
at higher values of 
${\bf |q|}$, corresponding to beam energies larger than $\sim 1$ \ GeV, 
the description of the final states $|X\rangle $ 
in terms of nonrelativistic nucleons is no longer possible. 
Calculations of $\WmunuA$ in this regime require a set of simplifying 
assumptions, allowing one to take into account the relativistic motion of 
final state particles carrying momenta $\sim {\bf q}$ as well as the occurrence 
of inelastic processes, leading to the 
appearance of hadrons other than protons and neutrons.

\subsection{The impulse approximation}

The main assumptions underlying the impulse approximation (IA) scheme are that i)
as the spatial resolution of a probe delivering momentum ${\bf q}$ 
is $\sim 1/|{\bf q}|$, at large enough $|{\bf q}|$ the target nucleus is seen 
by the probe as a collection of individual nucleons and ii) 
the particles produced at the interaction vertex and the recoiling (${\rm A}-1$)-nucleon 
system evolve indipendently of one another, which amounts to neglecting {\it both} 
statistical 
correlations due to Pauli blocking and dynamical Final State Interactions (FSI), i.e. 
rescattering processes driven by strong interactions.

\begin{figure}[hbt]
\centerline%
{\psfig{figure=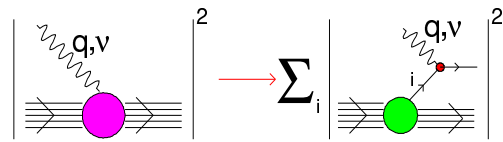,angle=00,width=6.50cm,height=1.75cm}}
\caption{\small (Color online)
Pictorial representation of the IA scheme, in which
the nuclear cross section is replaced by the incoherent sum of
cross sections describing scattering off individual bound nucleons, the
recoiling $({\rm A}-1)$-nucleon system acting as a spectator.}
\label{fig:1}
\end{figure}

In the IA regime the scattering process off a nuclear target reduces to the
incoherent sum of elementary processes involving only one nucleon, as
schematically illustrated in Fig. \ref{fig:1}. 

Within this picture, the nuclear current can be written as a sum of one-body 
currents
\beq
J^\mu \rightarrow \sum_i j_i^\mu \ ,
\label{currIA}
\eeq
while the final state reduces to the direct product of the hadronic state 
produced at the electromagnetic vertex, carrying momentum 
${\bf p}_x$ and the $(A-1)$-nucleon residual 
system, carrying momentum ${\bf p}_{\cal R}= {\bf q}-{\bf p}_x$ (for simplicity, 
we omit spin indices)
\beq
|X\rangle \rightarrow |x,{\bf p}_x\rangle
\otimes |{\cal R},{\bf p_{\cal R}}\rangle \ .
\label{resIA}
\eeq
Using Eq. (\ref{resIA}) we can rewrite the sum in Eq. (\ref{e:hadrten}) replacing
\be
\nonumber
\sum_X | X \rangle \langle X | & \rightarrow & \sum_{x} 
\int d^3p_x  | x,{\bf p}_x \rangle \langle {\bf p}_x,x | \\
& \times & \sum_{{\cal R}} d^3p_{{\cal R}} 
| {\cal R}, {\bf p}_{{\cal R}} \rangle \langle {\bf p}_{{\cal R}}, {\cal R} | \ .
\label{sumn}
\ee
Substitution of Eqs. (\ref{currIA})-(\ref{sumn}) into Eq. (\ref{e:hadrten}) and
insertion of a complete set of free nucleon states, satisfying 
\beq
\int d^3p\  | {\rm N},  {\bf p}\rangle\langle {\bf p},  {\rm N} |=I \ ,
\eeq
results in the factorization of the current matrix element 
\be
\nonumber
\langle 0 | J^\mu | X\rangle & = & \frac{m}{\sqrt{{\bf p}_{{\cal R}}^2 + m^2}}
 \langle 0 | {\cal R}, {\bf p}_{{\cal R}} ; {\rm N},-{\bf p}_{{\cal R}} \rangle \\ 
& \times & \sum_i \langle -{\bf p}_{\cal R},N | j^\mu_i | x,{\bf p}_x \rangle \ ,
\ee
leading to
\be
\nonumber
\WmunuA & = & \sum_{x,{\cal R}} \int d^3p_{{\cal R}}\ d^3p_x 
| \langle 0 | {\cal R},{\bf p}_{{\cal R}};{\rm N},-{\bf p}_{{\cal R}} \rangle |^2 
\  \frac{m}{ E_{{\bf p}_{\cal R}} } \\
\label{hadrtenIA}
& \times &  
\sum_i \ 
\langle -{\bf p}_{{\cal R}},{\rm N}| j^\mu_i | x,{\bf p}_x \rangle
\langle {\bf p}_x ,x | j^\nu_i | {\rm N},-{\bf p}_{{\cal R}} \rangle \\
\nonumber
& \times & 
\delta^{(3)}({\bf q}-{\bf p}_{{\cal R}}-{\bf p}_x)
\delta(\nu+E_0-E_{\cal R}-E_x),
\ee
where $E_{{\bf p}_{\cal R}} = 
\sqrt{|{\bf p}_{{\cal R}}|^2 + m^2}$. Finally, using the identity
\be
\nonumber
\delta(\nu+E_0-E_{\cal R}-E_x) & = & \int  
dE \delta(E-m+E_0-E_{\cal R}) \\
& \times & \delta (\nu-E+m-E_x) \ ,
\label{deltaIA}
\ee
and defining the target spectral function as \cite{foot1}
\be
\nonumber
P({\bf p}, E) & = & \sum_{\cal R}
|\langle 0|{\cal R},-{\bf p};{\rm N},{\bf p} \rangle |^2  \\
 & &  \ \ \ \ \ \times \delta(E-m+E_0-E_{\cal R}) \ ,
\label{specfunIA}
\ee
we can rewrite Eq. (\ref{e:hadrten}) in the form 
\be
\nonumber
\WmunuA({\bf q},\nu) & = & \sum_i
\int d^3p\ dE\ w_i^{\mu\nu}({\widetilde q})  \\
&  & \ \ \ \ \ \ \ \ \ \ \times
\left(\frac{m}{E_{{\bf p}}}\right)P({\bf p}, E) \ ,
\label{hadrten2}
\ee
with $E_{{\bf p}} = \sqrt{|{\bf p}^2|+m^2}$ and
\be
\nonumber
w_i^{\mu\nu} & = & \sum_x \langle {\bf p},{\rm N}| j^\mu_i | x,{\bf p}+{\bf q} \rangle
\langle {\bf p}+{\bf q},x | j^\nu_i | {\rm N},{\bf p} \rangle  \\
& \times & \delta({\widetilde \nu} + \sqrt{{\bf p}^2 + m^2} - E_x )\ .
\label{nucl:tens}
\ee
The quantity defined in the above equation is the tensor describing electromagnetic 
interactions of the $i$-th nucleon {\it in free space}. Hence, Eq. (\ref{nucl:tens}) 
shows that in the IA scheme the effect 
of nuclear binding of the struck nucleon is accounted for by the replacement
\beq
q \equiv (\nu,{\bf q}) \rightarrow {\widetilde q} \equiv ({\widetilde \nu},{\bf q}) \ ,
\label{def:qt}
\eeq
with (see Eqs. (\ref{hadrtenIA}) and (\ref{specfunIA}))
\be
\nonumber
{\widetilde \nu} & = & E_x - \sqrt{{\bf p}^2 + m^2} \\ 
\nonumber
       & = & \nu + E_0 - E_{\cal R} - \sqrt{{\bf p}^2 + m^2} \\
       & = & \nu - E  + m - \sqrt{{\bf p}^2 + m^2} \ ,
\label{def:nut}
\ee
in the argument of $w_i^{\mu\nu}$. This procedure essentially amounts to assuming 
that:
i) a fraction $\delta \nu$ of the energy transfer 
 goes into excitation energy of the spectator system and ii) the elementary 
scattering process can be described as if it took place in free space with 
energy transfer ${\widetilde \nu} = \nu - \delta \nu$. This interpretation 
emerges most naturally
in the $|{\bf p}| \ll m$ limit, in which Eq. (\ref{def:nut}) yields $\delta \nu = E$. 

Collecting together all the above results we can finally 
rewrite the doubly differential nuclear cross section in the form
\be
\nonumber
& & \frac{d\sigma_{IA}}{d\Omega_{e^\prime} dE_{e^\prime}} = \int \ d^3p \ dE \ 
P({\bf p},E)\ \left[ Z \frac{d\sigma_{ep}}{d\Omega_{e^\prime} dE_{e^\prime}}
 \right. \\
 &  & \ \ \ \ + 
\left. (A-Z) \frac{d\sigma_{en}}{d\Omega_{e^\prime} dE_{e^\prime}} 
\right] \delta(\nu-E+m-E_x) ,
\label{csIA}
\ee
where $d\sigma_{eN}/d\Omega_{e^\prime} dE_{e^\prime}$ ($N \equiv n,p$ denotes 
a proton or a neutron) is the cross section 
describing the elementary scattering process 
\beq
e(k) \ + \ N(p) \rightarrow e^\prime(k^\prime) 
+ x(p + {\widetilde q}) \ ,
\eeq
given by
\beq
\frac{d\sigma_{eN}}{d\Omega_{e^\prime} dE_{e^\prime}} = \frac{\alpha^2}{Q^4}
\frac{E_e^\prime}{E_e}  \  \frac{m}{E_{\bf p}} \Lmunu w^{\mu\nu}_N \ ,
\eeq
stripped of both the flux factor and the energy conserving $\delta$-function. 

\subsection{The nuclear spectral function}

In NMBT the nucleus is seen as a system of A nucleons whose dynamics are 
described by the nonrelativistic hamiltonian
\beq
H_A = \sum_{i=1}^{A} \frac{{\bf p}_i^2}{2m} + \sum_{j>i=1}^{A} v_{ij}
 + \sum_{k>j>i=1}^A V_{ijk} \ ,
\label{H:A}
\eeq
where ${\bf p}_i$ is the momentum of the $i$-th nucleon, while
 $v_{ij}$ and $V_{ijk}$ are
two- and three-nucleon interaction potentials, respectively. 

The two-nucleon potential, that reduces to
the Yukawa one-pion-exchange potential at large internucleon distance, is
obtained from an accurate fit to the available data on the
two-nucleon system, i.e. deuteron properties and $\sim$ 4000 NN 
scattering phase shifts at energies up to the pion production
threshold \cite{WSS}. The additional three-body term
$V_{ijk}$ has to be included in order to account for the binding energies of the
three-nucleon bound states \cite{PPCPW}.

The many-body Schr\"odinger equation associated with the Hamiltonian
of Eq. (\ref{H:A}) can be solved exactly, using stochastic methods,
for nuclei with mass number $A \leq 10$. The energies of the ground and low-lying 
excited states are in excellent agreement with the experimental
data \cite{WP}. Accurate calculations can also be carried out for uniform
nucleon matter, exploiting translational invariace and using either
a variational approach based on cluster expansion and chain summation 
techniques \cite{AP}, or G-matrix perturbation theory \cite{BGLS2000}.

Nonrelativistic NMBT provides a fully consistent computational 
framework that has been employed to obtain the spectral functions of 
the few-nucleon systems, having A$=3$ \cite{dieperink,cps,sauer} 
and 4 \cite{ciofi4,morita,bp}, 
as well as of nuclear matter, i.e. in the limit A $\rightarrow \infty$ 
with Z=A/2 \cite{bff,pkebbg}. Calculations based on 
G-matrix perturbation theory have also been carried out 
for oxygen \cite{geurts16,polls16}.

The spectral functions of different nuclei, ranging from Carbon to Gold, 
have been modeled using the
Local Density Approximation (LDA) \cite{bffs}, in which the experimental
information obtained from nucleon knock-out measurements is combined 
with the results of theoretical calculations of the nuclear matter 
$P({\bf p},E)$ at different densities \cite{bffs}.

Nucleon removal from shell model states 
has been extensively studied by coincidence $(e,e^\prime p)$ experiments 
(see, e.g., Ref. \cite{book}). The corresponding measured spectral function 
is usually written in the factorized form 
\beq
P_{MF}({\bf p},E) = \sum_n Z_n\ |\phi_n({\bf p})|^2 F_n(E-E_n) \ ,
\label{P:MF}
\eeq
where $\phi_{n}({\bf p})$ is the momentum-space wave function of the 
single particle shell mode state $n$, whose energy width is described 
by the function $F_n(E-E_n)$. The normalization of  the $n$-th state is 
given by the so called spectroscopic factor $Z_n < 1$, and the sum in 
Eq. (\ref{P:MF}) is extended to 
all occupied states of the Fermi sea. Hence, $P_{MF}({\bf p},E)$ vanishes at 
$|{\bf p}|$ larger than the 
Fermi momentum $p_F \sim 250$ MeV.
Note that in absence of NN correlations $F_n(E-E_n)$ shrinks to 
a $\delta$-function, $Z_n \equiv 1$ and 
Eq. (\ref{P:MF}) can be identified with the full spectral function.

Strong dynamical NN correlations give rise
to virtual scattering processes leading to the excitation of the participating
nucleons to states of energy larger than the Fermi energy, thus depleting
the shell model states within the Fermi sea. As a consequence, the spectral function
associated with nucleons belonging to correlated pairs extends to 
the region of $|{\bf p}| \gg p_F$ {\it and} $E \gg e_F$, where $e_F$ denotes
the Fermi energy, typically $\lsim 30$ MeV. 

The correlation contribution to $P({\bf p},E)$ of uniform nuclear matter 
has been calculated by Benhar {\it et al} for a wide range of density values \cite{bffs}.
Within the LDA scheme, the results of Ref. \cite{bffs} can be used to obtain the corresponding 
quantity for a finite nucleus of mass number $A$ from 
\beq
P_{corr}({\bf p},E) = \int d^3r\ \rho_A({\bf r})
P^{NM}_{corr}({\bf p},E;\rho = \rho_A({\bf r})) \ ,
\label{P:corr}
\eeq
where $\rho_A({\bf r})$ is the nuclear density distribution and
$P^{NM}_{corr}({\bf p},E;\rho)$ is the correlation component of
the spectral function of uniform nuclear matter at density $\rho$.

Finally, the full LDA nuclear spectral function can be written
\beq
P_{LDA}({\bf p},E) = P_{MF}({\bf p},E) + P_{corr}({\bf p},E) \ ,
\label{P:LDA}
\eeq
the spectroscopic factors $Z_n$ of Eq. (\ref{P:MF}) being constrained 
by the normalization requirement
\beq
\int d^3p\ dE\ P_{LDA}({\bf p},E) = 1\ .
\label{P:norm}
\eeq

\begin{figure}[hbt]
\centerline{\psfig{figure=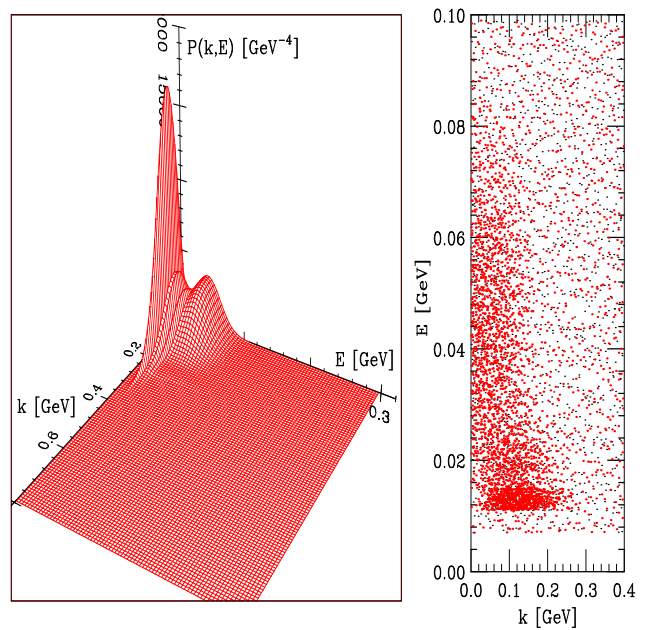
,angle=00,width=8.40cm,height=8.40cm}}
\caption{\small (Color online) 
Three-dimensional plot (left panel) and scatter plot 
(right panel) of the oxygen spectral function obtained using the LDA
approximation described in the text.}
\label{pke:O}
\end{figure}

The LDA spectral function of $^{16}O$ obtained combining the nuclear matter 
results of Ref. \cite{bffs} and the Saclay $(e,e^\prime p)$ data \cite{eep16O}
is shown in Fig. \ref{pke:O}. The shell model contribution $P_{MF}({\bf p},E)$ 
accounts for $\sim$ 80 \% of its normalization, whereas the remaining $\sim$ 20 \% 
of the strength, accounted for by $P_{corr}({\bf p},E)$, is located at high 
momentum ($|{\bf p}| \gg p_F$) {\it and} large removal energy ($E \gg e_F$). 
It has to be emphasized 
that large $E$ and large ${\bf p}$ are strongly correlated. For example, 
$\sim$ 50 \% of the strength at $|{\bf p}|$ = 320 MeV is located at 
$E >$ 80 MeV.

The LDA scheme rests on the premise that short range nuclear dynamics is
unaffected by surface and shell effects. The validity of this assumption
is confirmed by theoretical calculations of the nucleon momentum 
distribution, defined as
\be
\label{def1:nk}
n({\bf p}) & = & \int dE\ P({\bf p},E) \\
           & = & \langle 0 | a^\dagger_{\bf p} a_{\bf p} | 0 \rangle \ ,
\label{def2:nk}
\ee
where $a^\dagger_{\bf p}$ and $a_{\bf p}$ denote the creation and annihilation 
operators of a nucleon of momentum ${\bf p}$. The results clearly show that 
 for A$\ge 4$ the quantity $n({\bf p})/A$ becomes nearly independent of
$A$ in the region of large $|{\bf p}|$ ($\gsim 300$ MeV), where NN correlations 
dominate (see, e.g., Ref. \cite{rmp}).

\begin{figure}[hbt]
\centerline{\psfig{figure=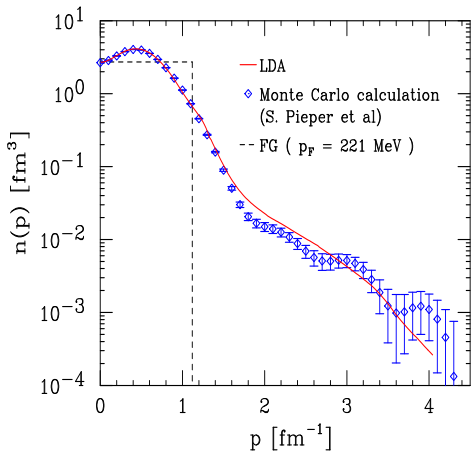
,angle=00,width=6.25cm,height=6.0cm}}
\caption{\small (Color online) 
Momentum distribution of nucleons in the oxygen ground state.
Solid line: LDA approximation. Dashed line: FG model with Fermi momentum
$p_F = 221$ MeV. 
Diamonds: Monte Carlo calculation carried out by S.C. Pieper \protect\cite{steve} 
using the 
wave function of Ref. \protect\cite{16Owf}.}
\label{nk:O}
\end{figure}

In Fig. \ref{nk:O} the nucleon momentum distribution 
of $^{16}$O, obtained from Eq. (\ref{def1:nk}) using the LDA spectral function 
of Fig. \ref{pke:O}, is compared to the one resulting from a 
Monte Carlo calculation \cite{steve}, carried out using the definition of 
Eq. (\ref{def2:nk}) and a highly realistic many-body wave function \cite{16Owf}. 
For reference, the FG model 
momemtum distribution corresponding to Fermi momentum $p_F$ = 221 MeV is also 
shown by the dashed line. It clearly appears that the $n({\bf p})$ obtained
from the spectral function is close to that of Ref.\cite{steve}, while the FG
distribution exhibits a completely different behaviour.
 
A direct measuremet of the correlation component of the 
spectral function of $^{12}C$, obtained measuring the
$(e,e^\prime p)$ cross section at missing momentum and energy up to $\sim$ 800 MeV
and $\sim 200$ MeV, respectively, has been recently carried out at Jefferson Lab by the
E97-006 Collaboration \cite{E97-006}. The data resulting from the preliminary
analysis appear to be consistent with the theoretical predictions based 
on LDA.

\subsection{Final state interactions}

The occurrence of strong FSI in electron nucleus scattering has long been
experimentally established. The results of a number of $(e,e^\prime p)$ measurements
covering the kinematical domain corresponding to $0.5 \lsim Q^2 \lsim 8.0$ GeV$^2$ 
 \cite{garino92,oneill95,abbott98,garrow02}, clearly show that the flux 
of outgoing protons is strongly suppressed, with respect to the IA predictions. 
The observed attenuation ranges from 20-40 \% in Carbon to 50-70 \% in Gold.

The inclusive $(e,e^\prime)$ cross section, being only sensitive to rescattering 
processes taking place within a distance $\sim 1/|{\bf q}|$ of the 
electromagnetic vertex, is obviously much less sensitive to FSI than the coincidence 
$(e,e^\prime p)$ cross section. The latter is in fact affected by rescatterings 
occurring over the distance $\sim R_A$, $R_A$ being the nuclear radius, travelled by the 
struck particle on its way out of the target. 
However, FSI effects become appreciable, indeed dominant, 
in the low $\nu$ region, where the inclusive cross section is most sensitive to 
the high momentum and high removal energy tails of the nuclear spectral function.

In quasi-elastic inclusive processes FSI produce two effects: i) an energy shift of 
the cross section,
due to the fact that the struck nucleon feels the mean field generated by the
spectator particles and ii) a redistribution of the strength, leading to the quenching
of the peak and the enhancement of the tails, to be ascribed to the occurrence
of NN rescattering processes that couple the one particle-one hole final state to
more complicated $n$ particle-$n$ holes configurations.

Early attempts to include FSI effects were based on the 
optical potential model \cite{horikawa80}. However, while providing a computationally 
practical scheme to account for the loss of flux in the one-nucleon removal channel,
this model relies on the mean field picture of the nucleus, and does not include the 
effect of dynamical NN correlations.

A different approach, based on NMBT and a generalization of Glauber theory
of high energy proton scattering \cite{glauber59} has been proposed by
Benhar et al. \cite{gofsix} in the early 90s. This treatment of FSI, generally 
referred to as
Correlated Glauber Approximation (CGA) rests on the assumptions that
i) the struck nucleon moves along a straight trajectory with constant velocity
(eikonal approximation), and ii) the spectator nucleons are seen by the
struck particle as a collection of fixed scattering centers
(frozen approximation).

Under the above assumptions the expectation value of the propagator of the
struck nucleon in the target ground state can be written in the factorized form
\beq
U_{{\bf p}+{\bf q}}(t) = U^0_{{\bf p}+{\bf q}}(t)
{\bar U}^{FSI}_{{\bf p}+{\bf q}}(t)\ ,
\eeq
where $U^0_{{\bf p}+{\bf q}}$ is the free space propagator, while FSI
effects are described by the quantity ($R \equiv ({\bf r}_1, \ldots, {\bf r}_A)$ 
specifies the target configuration)
\beq 
{\bar U}^{FSI}_{{\bf p}+{\bf q}}(t) = \langle 0 | 
U^{FSI}_{{\bf p}+{\bf q}}(R;t) | 0 \rangle \ ,
\label{eik:prop0}
\eeq
with
\beq
U^{FSI}_{{\bf p}+{\bf q}}(R;t) = \frac{1}{A} \sum_{i=1}^A
{\rm e}^{i \sum_{j \neq i} \int_0^t dt^\prime
w_{{\bf p}+{\bf q}}(|{\bf r}_{ij} + {\bf v}t^\prime |) } \ .
\label{eik:prop}
\eeq
In Eq. (\ref{eik:prop}) ${\bf r}_{ij}={\bf r}_{i}-{\bf r}_{j}$ and
$w_{{\bf p}+{\bf q}}(|{\bf r}|)$ is the coordinate-space NN scattering
t-matrix at incident momentum ${\bf p}+{\bf q}$, usually parametrized in terms of 
total cross
section, slope and real to imaginary part ratio. 
At large $|{\bf q}|$, ${\bf p}+{\bf q} \approx {\bf q}$ and the eikonal 
propagator of Eq. (\ref{eik:prop0}) becomes a function
of $t$ and the momentum transfer only. 

Note that $U^{FSI}_{{\bf q}}(R;t)$ is simply related to the nuclear
transparency $T_{A}$, measured in coincidence $(e,e^\prime p)$ experiments
\cite{garino92,oneill95,abbott98,garrow02}, through
\beq
T_{A} = \lim_{t \rightarrow \infty} \langle 0 | |U^{FSI}_{{\bf q}}(R;t) |^2 | 0 \rangle \ .
\eeq

\begin{figure}[hbt]
\centerline{\psfig{figure=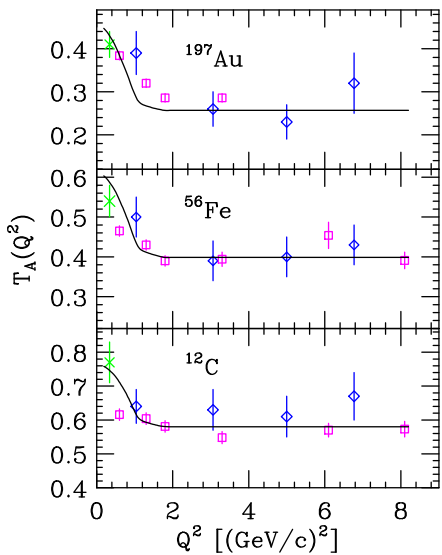,angle=00,width=5.75cm,height=7.25cm}}
\caption{\small (Color online)
$Q^2$-dependence of the transparency of Carbon, Iron and Gold \protect\cite{daniela}, 
calculated using LDA and the approach of ref.\protect\cite{gofsix}. The data points 
are taken from
Refs.\protect\cite{garino92} (crosses), \cite{oneill95} (diamonds) and
\cite{abbott98,garrow02} (squares). Note that, in absence of FSI, 
$T_A(Q^2) \equiv 1$.
}
\label{transp}
\end{figure}

The results displayed in Fig. \ref{transp} \cite{daniela} show that both the magnitude 
and 
the $A$- and $Q^2$-dependence of the transparencies of Carbon, Iron and Gold 
obtained from the approach of Ref. \cite{gofsix} and LDA are in good agreement 
with the experimental data. Note that at low $Q^2$ FSI lead to a $\sim$ 20 (40) \% effect 
in Carbon (Iron). Neglecting this effect, i.e. setting $T_A(Q^2) \equiv 1$, would
be utterly incompatible with the data.

From Eqs. (\ref{eik:prop0}) and  (\ref{eik:prop}) it follows that within the approach 
of Ref. \cite{gofsix} the energy 
shift and the redistribution of the inclusive strength are driven by the real and 
the imaginary part of the NN scattering amplitude, respectively. At large 
${\bf q}$ the imaginary part of $w_{{\bf q}}$, corresponding to the real part of 
$U^{FSI}_{{\bf q}}$, is dominant. 
Neglecting the contribution of the real part of $w_{{\bf q}}$ altogether, 
the CGA quasi-elastic inclusive cross section can be written as
a convolution integral, involving the cross section evaluated within the
 IA, i.e. in absence of FSI, and a folding function embodying FSI effects:
\beq
\frac{d\sigma}{d\Omega_{e^\prime} d\nu} = \int d\nu^\prime \
f_{{\bf q}}(\nu - \nu^\prime)\ 
\left( \frac{d\sigma}{d\Omega_{e^\prime} d\nu^\prime} \right)_{IA} \ ,
\label{sigma:FSI}
\eeq
the folding function $f_{{\bf q}}(\nu)$ being defined as
\beq
f_{{\bf q}}(\nu) = \delta(\nu) \sqrt{T_{A}} \\
 + \int \frac{dt}{2 \pi}\ {\rm e}^{i \nu t}
\left[ U^{FSI}_{{\bf q}}(t) - \sqrt{T_{A}} \right]\ .
\label{ff}
\eeq
The above equations clearly show that the strength of FSI is measured by
both $T_{A}$ and the width of the folding function. In absence
of FSI
$U^{FSI}_{{\bf q}}(R;t) \equiv 1$, implying in turn
$T_{A}=1$ and $f_{{\bf q}}(\nu) = \delta(\nu)$.

Dynamical NN correlations stronlgy affect the shape of the
folding function of Eq. (\ref{ff}). Due to the strong repulsive core of the NN
force, the joint probability of finding two nucleons at positions ${\bf r}_i$
 and ${\bf r}_j$, driving the occurrence of rescattering processes in the final 
state, is strongly suppressed at $|{\bf r}_i - {\bf r}_j| \lsim 1$ fm. 
As a consequence, inclusion of correlation effects within the framework 
of NMBT leads to a strong quenching of FSI effects, with respect to the 
predictions of the independent particle model.

In principle, the real part of the NN scattering ampliutde 
can be explicitely included in Eq. (\ref{eik:prop}) and treated on the same footing 
as the imaginary part. However, its effect turns out to be appreaciable only at 
$t \sim 0$, when the attenuation 
produced by the imaginary part is weak. The results of numerical calculations 
show that an approximate treatment based on the use of a time independent 
optical potential is indeed adequate to describe the energy shift produced 
by the real part of $w_{{\bf q}}$ \cite{bffs}, whose size of $\sim 10$ MeV
is to be compared to a typical electron energy loss of few hundreds MeV.

\section {Comparison to electron scattering data}
\label{eA-results}

We have employed the formalism described in the previuos Sections to compute
the inclusive electron scattering cross section off oxygen at 
$0.2 \lsim Q^2 \lsim 0.6 $ GeV$^2$. 

The IA cross section has been obtained using the LDA spectral function shown 
in Fig. \ref{pke:O} and the nucleon tensor defined by Eq. (\ref{nucl:tens}),
that can be written as
\be 
\nonumber
w_N^{\mu\nu} & = & w^N_1 \left( -g^{\mu\nu} + \frac{\qt^\mu \qt^\nu}{\qt^2} 
  \right)  \\
 & + & \frac{w^N_2}{m^2} \left(p^\mu - \frac{(p \qt)}{\qt^2}q^\mu \right)
                   \left(p^\nu - \frac{(p \qt)}{\qt^2}q^\nu \right) \ ,
\ee
where $p\equiv (E_{\bf p},{\bf k})$ and the off-shell four momentum transfer 
$\qt$ is defined by Eqs. (\ref{def:qt}) and (\ref{def:nut}). The two structure 
functions $w^N_1$ and $w^N_2$
are extracted from electron-proton and electron-deuteron scattering data. In the case of 
quasi-elastic scattering they are simply related to the electric 
and magnetic nucleon form factors, $G_{E_N}$ and $G_{M_N}$, through
\beq
w^N_1 = -\frac{\qt^2}{4m^2}\ \delta\left({\widetilde \nu} + \frac{\qt^2}{2m} \right) 
\ G_{M_N}^2 \ ,
\eeq
\be
\nonumber
w^N_2 & = & \frac{1}{1 - \qt^2/4 m^2} \ 
\delta\left({\widetilde \nu} + \frac{\qt^2}{2m} \right) \\
  & & \ \ \ \ \ \ \ \ \ \ \ \ \ \ \ 
\times \left( G_{E_N}^2 - \frac{\qt^2}{4m^2} G_{M_N}^2 \right)\ .
\ee
Numerical calculations have been carried out using 
the H\"ohler-Brash parameterization of the form factors \cite{Hohler76,Brash02}, 
resulting from a fit which includes the recent Jefferson Lab data \cite{Jones00}. 

In the kinematical region under discussion, inelastic processes, mainly 
quasi-free $\Delta$ resonance production, are also known to play a role. To include these 
contributions in the calculation of the inclusive cross section, we have adopted the 
Bodek and Ritchie parametrization of the proton and neutron structure functions \cite{br}, 
 covering both the resonance and deep inelastic region.

The folding functions describing the effect of NN rescattering in the final 
state have been computed from Eq. (\ref{ff}) with the eikonal propagator 
$U^{FSI}_{{\bf q}}(R;t)$ obtained using the parametrization of the NN scattering 
amplitude of Ref. \cite{oneillNN} and the medium modified NN cross sections of 
Ref. \cite{papi}. The integrations involved in Eq. (\ref{eik:prop0})
have been carried out using Monte Carlo configurations sampled from the probability
distribution associated with the oxygen ground state wave function of 
Ref. \cite{16Owf}. 

The effect of the real part of the NN scattering amplitude has been approximated 
including in the energy conserving $\delta$-function of Eq. (\ref{csIA}) the real 
part of the optical potential felt by a nucleon of momentum ${\bf p}+{\bf q}$ 
embedded in uniform nuclear matter at equilibrium density.

\begin{figure}[hbt]
\centerline%
{\psfig{figure=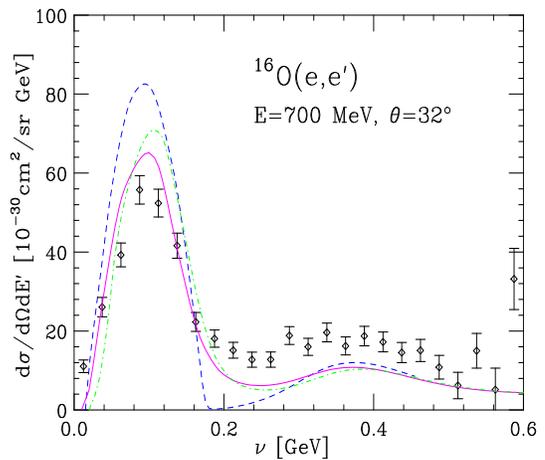,angle=00,width=7.0cm,height=6.0cm}}
\caption{\small (Color online) Cross section of the process $^{16}O(e,e^\prime)$
at beam energy 700 MeV and electron scattering angle 32$^\circ$. 
Solid line: full calculation, carried out within the approach described 
in Section \protect\ref{eA-formalism}. Dot-dash line: IA calculation, carried out
neglecting FSI effects. Dashed line: FG model with $p_F = 225$ MeV and 
$\epsilon = 25$ MeV. The experimental data are from Ref.\protect\cite{LNF}.}
\label{fig:ee1}
\end{figure}

\begin{figure}[hbt]
\centerline%
{\psfig{figure=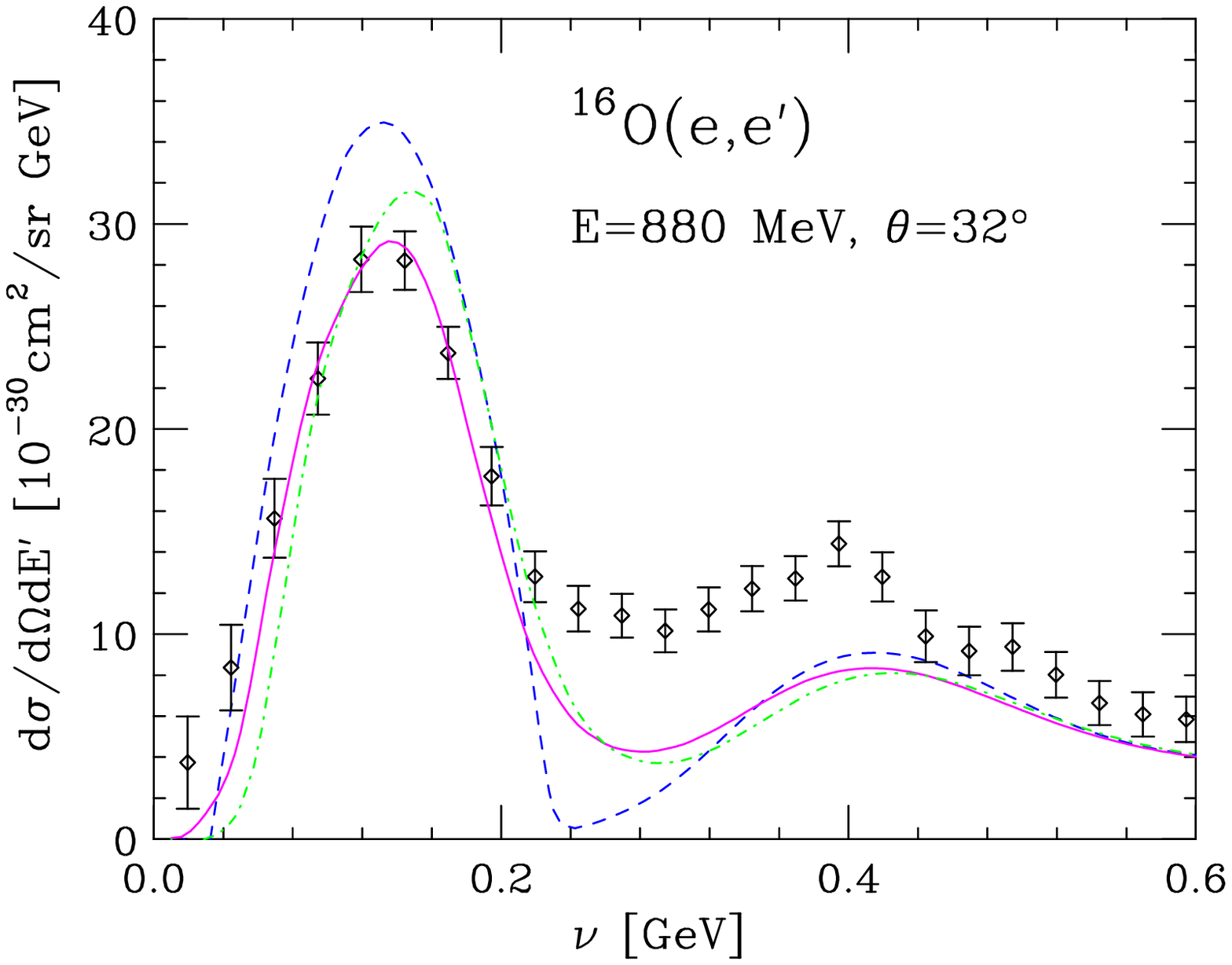,angle=00,width=7.0cm,height=6.0cm}}
\caption{\small (Color online)
Same as in Fig. \protect\ref{fig:ee1}, but for beam energy 880 MeV.}
\label{fig:ee2}
\end{figure}

\begin{figure}[hbt]
\centerline%
{\psfig{figure=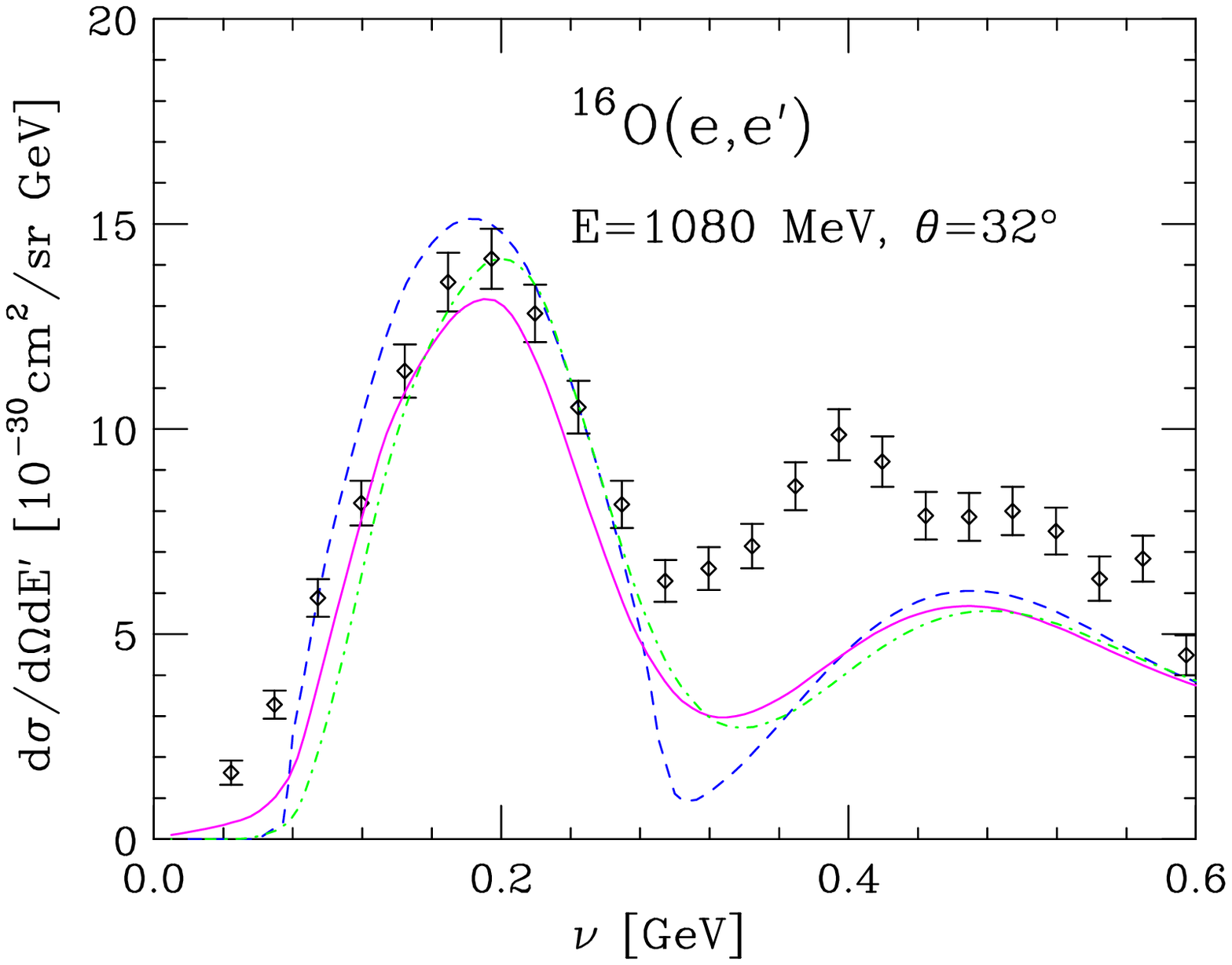,angle=00,width=7.0cm,height=6.0cm}}
\caption{\small (Color online) 
Same as in Fig. \protect\ref{fig:ee1}, but for beam energy 1080 MeV.}
\label{fig:ee3}
\end{figure}

\begin{figure}[hbt]
\centerline%
{\psfig{figure=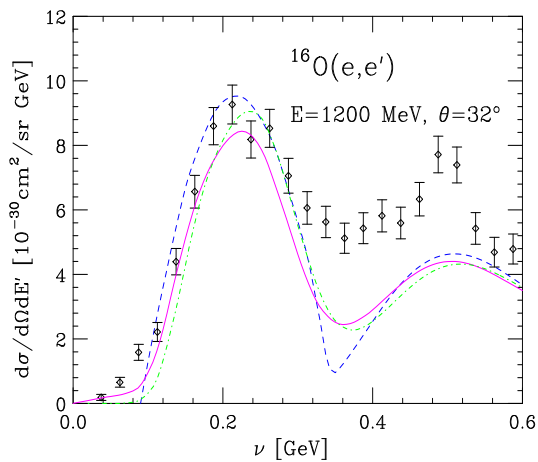,angle=00,width=7.0cm,height=6.0cm}}
\caption{\small (Color online) 
Same as in Fig. \protect\ref{fig:ee1}, but for beam energy 1200 MeV.}
\label{fig:ee4}
\end{figure}

In Figs. \ref{fig:ee1}-\ref{fig:ee4} the results of our calculations
are compared to the data of Ref. \cite{LNF}, corresponding to beam energies
700, 880, 1080 and 1200 MeV and electron scattering angle 32$^\circ$. For 
reference, the results of the FG model corresponding to Fermi 
momentum $p_F = 225$ MeV and average removal energy $\epsilon = 25$ MeV
are also shown.

Overall, the approach described in the previous Sections, {\it involving no
adjustable parameters}, provides a fairly accurate account of the measured
cross sections in the region of the quasi-free peak. On the other hand, the
FG model, while yielding a reasonable description 
at beam energies 1080 and 1200 MeV, largely overestimates the data at lower energies. 
The discrepancy at the top of the quasi-elastic peak turns out to be $\sim$ 25 \% 
and $\sim$ 50 \% at 880 and 700 MeV, respectively.

The results of NMBT and FG model also turn out to be sizably different
in the dip region, on the right hand side of the quasi-elastic peak, while
the discrepancies become less pronounced at the $\Delta$-production peak.
However, it clearly appears that, independent of the employed approach and beam 
energy, theoretical 
results significantly underestimate the data at energy transfer larger than the
pion production threshold.

\begin{figure}[hbt]
\centerline{\psfig{figure=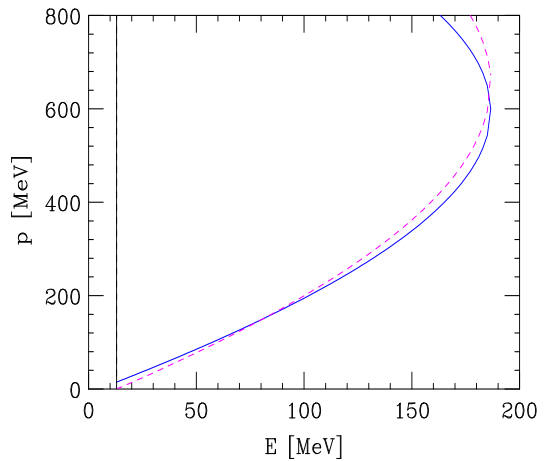
,angle=00,width=7.00cm,height=6.00cm}}
\caption{\small (Color online)
The solid and dashed lines enclose the integration regions in the 
$({\bf p},E)$ plane relevant to the calculation of the IA cross section at 
the top of the quasi-elastic and $\Delta$ production peak, respectively, for beam 
energy 1200 MeV and scattering angle 32$^\circ$.}
\label{nu:0}
\end{figure}

In view of the fact that the quasi-elastic peak is correctly reproduced
(within an accuracy of $\sim$ 10 \%), the failure of NMBT to reproduce the data 
at larger $\omega$ may be ascribed to deficiencies in the description of the 
elementary electron-nucleon cross section. In fact, as ilustrated 
in Fig. \ref{nu:0}, the calculation of the IA 
cross section at the quasi-elastic and $\Delta$ production peak involves 
integrations of $P({\bf p},E)$ extending over regions of the 
$({\bf p},E)$ plane almost exaclty overlapping one another. 

\begin{figure}[hbt]
\centerline{\psfig{figure=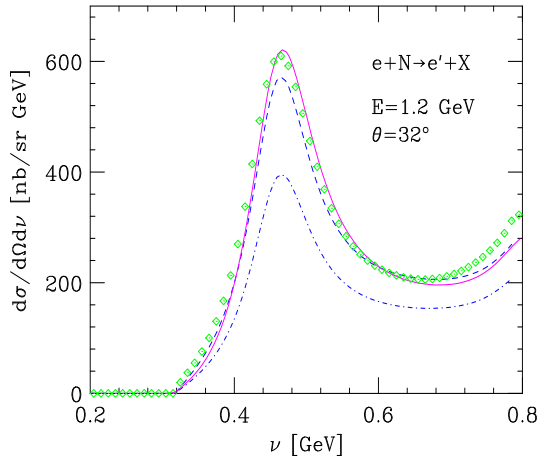
,angle=00,width=7.00cm,height=6.00cm}}
\caption{\small (Color online)
Cross section of the process $e+N \rightarrow e^\prime + X$ above pion 
production threshold, at beam energy 1200 MeV and scattering angle 32$^\circ$. 
Solid line: fit of Ref. \protect\cite{thia} for $ep$ scattering; dashed line:
fit of Ref. \protect\cite{br} for $ep$ scattering; diamonds: fit of 
Ref. \protect\cite{christy} for $ep$ scattering; 
dot-dash line: fit of Ref. \protect\cite{br} for $en$ scattering.  }
\label{eN:fit}
\end{figure}

To gauge the uncertainty associated with the description of the nucleon structure 
functions $w_1^N$ and $w_2^N$, we have compared the electron-proton cross sections 
obtained from the model of Ref. \cite{br} to the ones obtained from the $H_2$ model
of Ref. \cite{thia} and from a global fit including recent Jefferson Lab 
data \cite{christy}. The results of 
Fig. \ref{eN:fit} show that at $E_e=1200$ MeV and $\theta=32^\circ$ the 
discrepancy between the different models is not large, being $\sim$ 15 \% at 
the $\Delta$ production peak.  
It has to be noticed, however, that the models of Refs. \cite{br,thia,christy}
 have all been obtained fitting data taken at electron 
beam energies larger than 2 GeV, so that their use in the kinematical regime 
discussed in this work involves a degree of extrapolation.

\begin{figure}[hbt]
\centerline{\psfig{figure=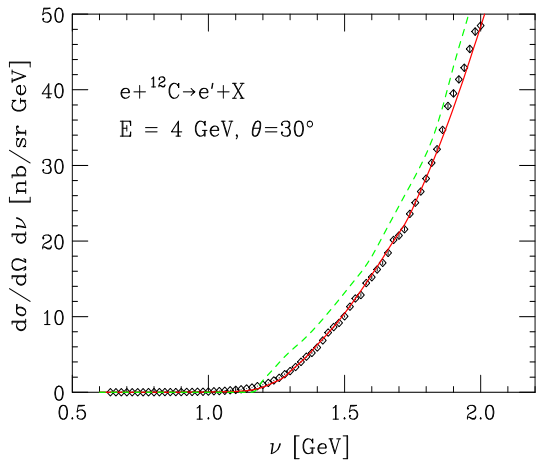
,angle=00,width=7.00cm,height=6.00cm}}
\caption{\small (Color online)
Cross section of the process $e+ ^{12}C \rightarrow e^\prime + X$ 
at beam energy 4 GeV and scattering angle 32$^\circ$.
The solid line has been obtained using the approach described
in this work, while the dashed line shows the results of the FG model
corresponding to $p_F = 221$ MeV and $\epsilon = 25$ MeV. The experimental 
data are from Ref. \protect\cite{E89-008}.}
\label{c4gev}
\end{figure}

On the other hand, the results obtained using the approach described in this 
paper and the nucleon structure functions of Ref. \cite{br}
are in excellent agreement with the measured $(e,e^\prime)$ cross sections
at beam energies of few GeV \cite{bffs}. As an example, in Fig. \ref{c4gev} we 
show a comparison between the calculated $^{12}C(e,e^\prime)$ cross section
and the Jefferson Lab data of Ref. \cite{E89-008} at $E_e=4$ GeV and
$\theta=30^\circ$. The corresponding FG result is also displayed, for reference.

\begin{figure}[hbt]
\centerline{\psfig{figure=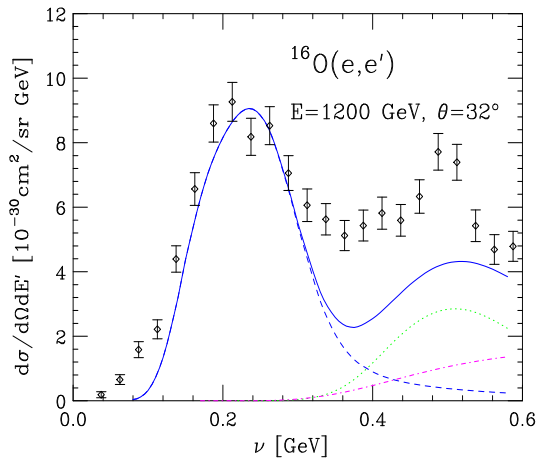
,angle=00,width=7.00cm,height=6.00cm}}
\caption{\small (Color online)
IA cross section of the process $^{16}O(e,e^\prime)$ at beam energy 1200 MeV and
scattering angle 32$^\circ$.
Dashed line: quasi-elastic; dots: quasi-free $\Delta$ production;
dashes: nonresonant background; solid line: total. The experimental data are
from Ref. \protect\cite{LNF}.}
\label{back}
\end{figure}

Fig. \ref{eN:fit} also shows the prediction of the Bodek and Ritchie fit for
the neutron cross section, which turns out to be much smaller than the proton one.
It is on account of this difference that we have chosen to adopt the fit of 
Ref. \cite{br}, as 
it  allows for a consistent inclusion of proton and neutron contributions, 
both resonant and nonresonsant, to the nuclear cross section. In this regard, 
it has to be pointed out that the nonresonant backround is not negligible.
As illustrated in Fig. \ref{back}, for beam energy 1200 MeV and scattering
angle 32$^\circ$ it provides $\sim$ 25 \% of the cross section
at energy transfer corresponding to the $\Delta$ peak.

The folding function described in Section \ref{eA-formalism} accounts for the FSI 
between a nucleon and the spectator system. For this reason, the results shown in 
Figs. \ref{fig:ee1}-\ref{fig:ee4} have been obtained folding with 
$f_{{\bf q}}(\nu)$ only the quasielastic component of the IA cross section.
Particles other than protons and neutrons, that can be produced at the electromagnetic 
vertex, also have FSI, but they are more difficult to describe. However, the inelastic
part of the IA cross section, being rather smooth, is unlikely to be strongly 
affected by FSI. To gauge the possible relevance of neglecting FSI in the inelastic 
channels we have computed the cross section at incident energy 880 MeV and 
scattering angle 32$^\circ$ folding the total IA result. Comparison with the
results displayed in Fig. \ref{fig:ee2} shows a difference of $\sim$ 0.5 \% at the
top of the $\Delta$-production peak.

\section{Charged current neutrino-nucleus cross section}
\label{nuA}

The Born approximation cross section of the weak charged current process
\beq
\nu_\ell + A \rightarrow \ell^- + X\ ,
\label{nu:process}
\eeq
can be written in the form (compare to Eq. (\ref{e:cross:section}))
\beq
\frac{d\sigma}{d\Omega_\ell dE_\ell} = \frac{G^2}{32 \pi^2}\
\frac{|{\bf k}^\prime|}{|{\bf k}|}\
 L_{\mu \nu} W^{\mu \nu}\ ,
\label{nu:cross:section}
\eeq
where $G=G_F \cos \theta_C$, $G_F$ and $\theta_C$ being Fermi's coupling constant and
Cabibbo's angle, $E_\ell$ is the energy of the final state
lepton and ${\bf k}$ and ${\bf k}^\prime$ are the neutrino and charged lepton
momenta, respectively. Compared to the corresponding quantities appearing in
Eq. (\ref{e:cross:section}), the tensors $L_{\mu \nu}$ and
$W^{\mu \nu}$ include additional terms resulting from the
presence of axial-vector components in the leptonic and hadronic
currents (see, e.g., Ref. \cite{walecka}).

Within the IA scheme, the cross section of Eq. (\ref{nu:cross:section}) can be cast
in a form similar to that obtained for the case of electron-nucleus scattering
(see Eq. (\ref{csIA})). Hence, its calculation requires the nuclear spectral function
and the tensor describing the weak charged current interaction of a free nucleon,
$w_N^{\mu\nu}$. In the case of quasi-elastic scattering, neglecting the contribution
associated with the pseudoscalar form factor $F_P$, the latter can be written
in terms of the nucleon Dirac and Pauli form factors $F_1$ and $F_2$, related to the
measured electric and magnetic form factors $G_{E}$ and $G_{M}$ through
\beq
F_{1} = \frac{1}{1 - q^2/4 m^2} \left( G_{E} - \frac{q^2}{4 m^2} G_{M} \right)
\eeq
\beq
F_{2} = \frac{1}{1 - q^2/4 m^2} \left( G_{M} - G_{E} \right) \ ,
\eeq
and the axial form factor $F_A$.

\begin{figure}[hbt]
\centerline{\psfig{figure=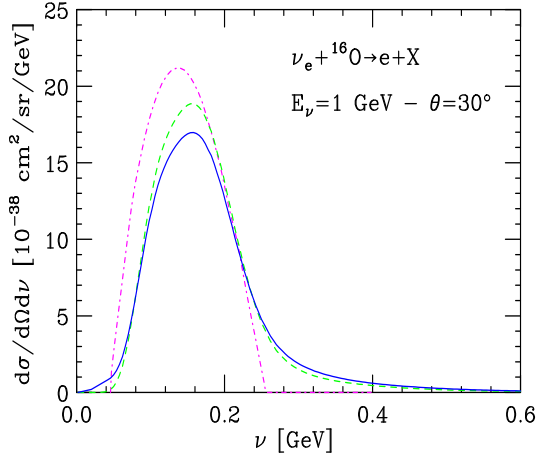
,angle=00,width=7.00cm,height=6.00cm}}
\caption{\small (Color online)
Differential cross section $d\sigma/d\Omega_e d\nu$
for neutrino energy $E_\nu = 1 $ GeV
and electron scattering angle $\theta_e = 30^\circ$.
The IA results are represented by the dashed line, while the solid line
corresponds to the full calculation, including the effects of FSI.
The dotted line shows the prediction of the FG model
with Fermi momentum $k_F = 225$ MeV and average separation energy
$\epsilon = 25$ MeV.}
\label{nu:1}
\end{figure}

Figure \ref{nu:1} shows the calculated cross section of the process $^{16}O(\nu_e,e)$,
corresponding to neutrino energy $E_\nu = 1 $ GeV and electron scattering
angle $\theta_e = 30^\circ$, plotted as a function of the energy transfer
$\nu = E_\nu - E_e$.
Numerical results have been obtained using the spectral function of Fig. \ref{pke:O}
and the dipole parametrization for the form factors, with an axial mass of 1.03 GeV.

Comparison between the solid and dashed lines shows that
the inclusion of FSI results in a sizable redistribution of the IA strength,
leading to a quenching of the quasi-elastic peak and to the enhancement of the tails.
For reference, we also show the cross section predicted by the FG
 model with Fermi momentum $p_F = 225$ MeV and average separation energy
$\epsilon = 25$ MeV. Nuclear dynamics, neglected in the oversimplified
picture in terms of noninteracting nucleons, clearly appears to play a relevant role.

\begin{figure}[hbt]
\centerline{\psfig{figure=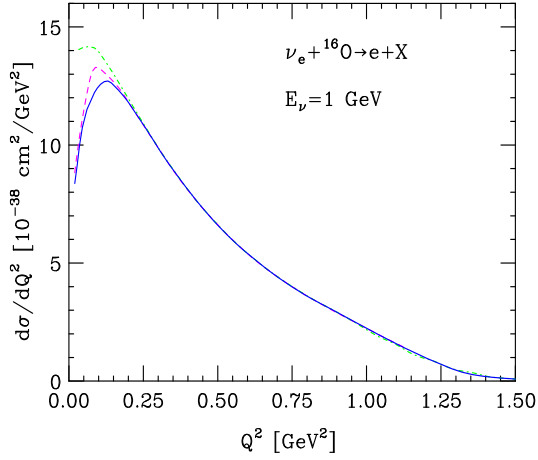
,angle=00,width=7.00cm,height=6.00cm}}
\caption{\small (Color online)
Differential cross section $d\sigma/dQ^2$
for neutrino energy $E= 1$ GeV. The dot-dash line shows the IA results,
while the solid and dashed lines have been obtained using the modified
spectral function ef Eq. (\protect\ref{pauli}), with and without inclusion
of FSI, respectively.
.}
\label{nu:2}
\end{figure}

It has to be pointed out that the approach described in
Section \ref{eA-formalism}, while including dynamical
correlations in the final state, does not take into account statistical
correlations, leading to Pauli blocking of the phase space available to the
knocked-out nucleon.

A rather crude prescription to estimate the effect of Pauli blocking amounts to
modifying the spectral function through the replacement
\beq
P({\bf p},E) \rightarrow P({\bf p},E)
\theta(|{\bf p} + {\bf q}| - {\overline p}_F)
\label{pauli}
\eeq
where ${\overline p}_F$ is the average nuclear Fermi momentum, defined as
\beq
{\overline p}_F = \int  d^3r\ \rho_A({\bf r}) p_F({\bf r}),
\label{local:kF}
\eeq
with $p_F({\bf r})=(3 \pi^2 \rho_A({\bf r})/2 )^{1/3}$, $\rho_A({\bf r})$ being the
nuclear density distribution. For oxygen, Eq. (\ref{local:kF}) yields
${\overline p}_F = 209$ MeV. Note that, unlike the spectral function, the
quantity defined in Eq. (\ref{pauli})
does not describe intrinsic properties of the target only, as it depends
explicitely on the momentum transfer.

\begin{figure}[ht]
\begin{center}
\includegraphics[scale=0.65]{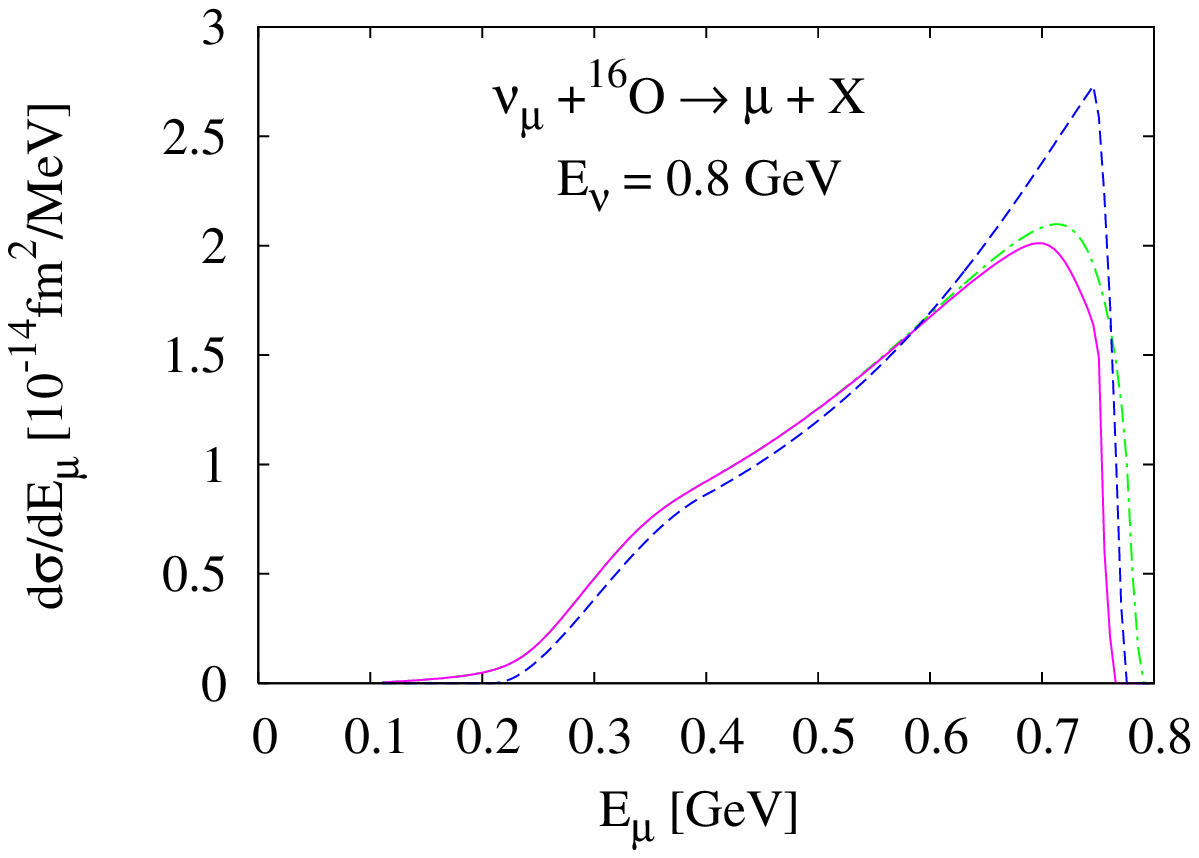}
\includegraphics[scale=0.65]{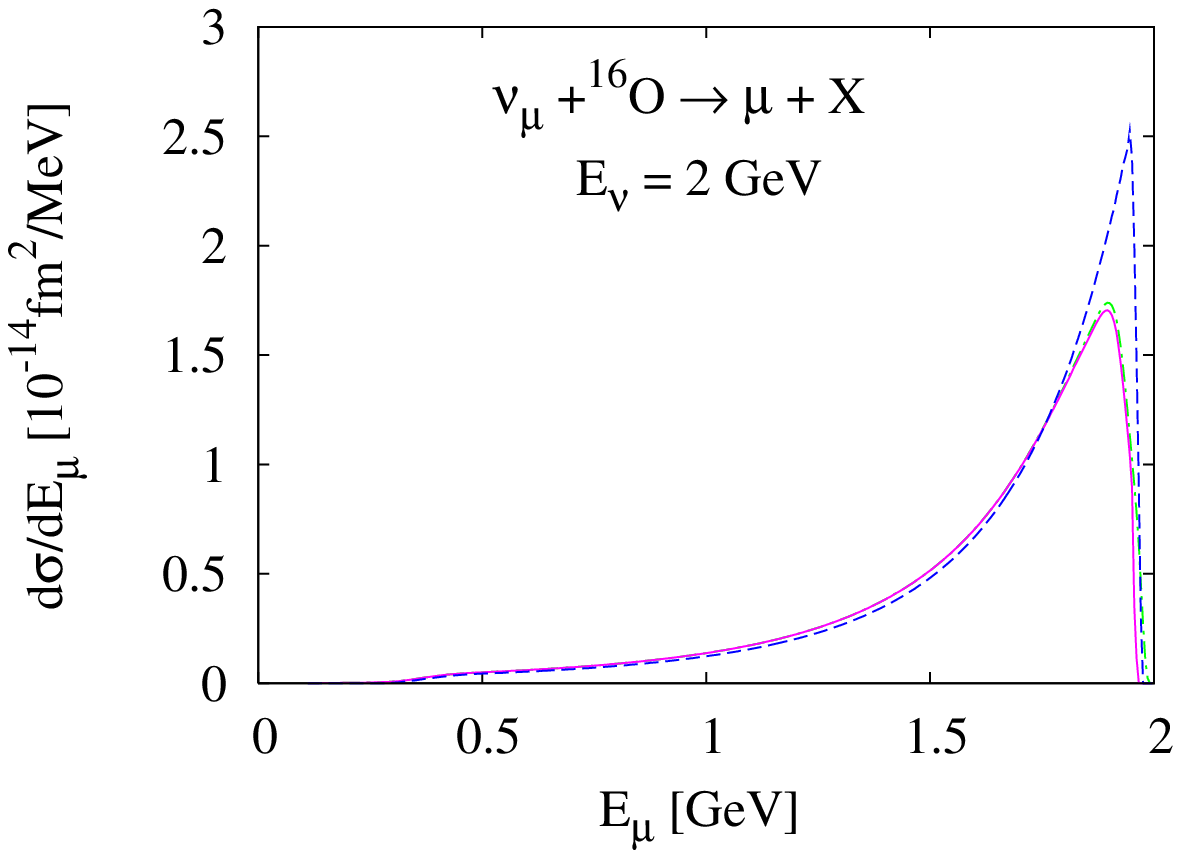}
\end{center}
\caption{
(Color online) Quasi-elastic differential cross section $d\sigma/dE_\mu$
as a function of the scattered energy $E_\mu$ for the neutrino energy $E =$ 0.8
and 2.0 GeV. The solid line shows IA calculation with Pauli
blocking as in Eq. (40), the dot-dash line IA calculation without Pauli
blocking, and the dashed line FG model.}
\label{del}
\end{figure}

The effect of Pauli blocking is hardly visible in the differential cross section
shown in Fig. \ref{nu:1}, as the kinematical setup corresponds to
$Q^2 > 0.2$ GeV$^2$ at the quasi-elastic peak. The same is true for the 
electron scattering cross sections discussed in the previous Section. 
On the other hand, this effect becomes very large at lower $Q^2$.

Figure \ref{nu:2} shows the calculated differential cross section $d\sigma/dQ^2$
for neutrino energy $E_\nu= 1$ GeV. The dashed and dot-dash lines correspond to the
IA results with and without inclusion of Pauli blocking, respectively. It
clearly appears that the effect of Fermi statistic in suppressing scattering
shows up at $Q^2 < 0.2$ GeV$^2$ and becomes very large at lower $Q^2$. The results of
the full calculation, in which dynamical FSI are also included, are displayed as a full line.
The results of Fig. \ref{nu:2} suggest that Pauli blocking and FSI may explain
the deficit of the measured cross section at low $Q^2$ with respect to the
predictions of Monte Carlo simulations \cite{Ishida}.

Figure \ref{del} shows the $\nu_\mu$-nucleus cross sections
as a function of the scattered muon energy, by comparing
the cross sections calculated by FG, and by the use of the spectral
function with and without Pauli blocking.  Figure \ref{del} shows
that FG yields a larger high-energy peak contribution than the other
two. This is {\it not} due to the Pauli blocking, but due to the
nuclear correlation effects in the spectral function: the muons
tend to be scattered with a higher energy. This effect should show
up in the forward angle cross section and  {\it may have a direct
effect on neutrino oscillation measurements}.

\section{Conclusions}
\label{conclusions}

We have employed an approach based on NMBT to compute the inclusive 
electron- and neutrino-nucleus scattering cross sections in the kinematical region 
corresponding to beam energy $\sim$ 1 GeV, relevant to many neutrino 
oscillation experiments.
Our calculations have been carried out within the IA scheme, using realistic 
spectral functions obtained from $(e,e^\prime p)$ data and theoretical calculations 
of uniform nuclear matter.

In the region of the quasi-elastic peak, the results of our calculations account
for the measured $^{16}O(e,e^\prime)$ cross sections at beam energies between
700 MeV and 1200 MeV and scattering angle 32$^\circ$ with an accuracy better
than 10 \%. It must be emphasized that the ability to yield 
quantitative predictions over a wide range of beam energies is critical 
to the analysis of neutrino experiments, in which the energy of the incident
neutrino is not known, and must be reconstructed from the kinematics of the 
outgoing lepton. 

In the region of quasi-free $\Delta$ production theoretical predictions 
significantly underestimate the data. Assuming the validity of the IA scheme, 
this problem appears to be mainly ascribable to uncertainties in the 
description of the nucleon structure functions in this kinematical 
regime. The upcoming electron-nucleus scattetring data in the resonance region 
from the Jefferson Lab E04-001 experiment \cite{thia2} will help to shed light 
on this issue.  At higher energies, i.e. in the region in which inelastic contributions
largely dominate, the calculated cross sections are in close agreement 
with the data.

Among the mechanisms not included in the IA picture, scattering processes 
in which the incoming lepton couples to meson exchange currents are not 
expected to produce large corrections to our results in the region of the
quasi-elastic peak. Numerical studies of the transverse response 
of uniform nuclear matter, carried out within 
NMBT \cite{fabro}, have shown that inclusion of two-body contributions to the 
nuclear electromagnetic current, arising from $\pi$ and $\rho$
meson exchange, leads to an enhancement that decreases as the
momentum transfer increases, and never exceeds 10 \% at $Q^2 < 0.25$ GeV$^2$.
On the other hand, the results of calculations of the tranverse response of 
the few-nucleon systems (for a review see Ref. \cite{rocco}) suggest that 
two-body current contributions may play a role in the dip region, at least 
for the lower values of the momentum transfer.

The second mechanism not included in the IA, Pauli blocking, while 
not appreciably affecting the lepton energy loss spectra, produces a large effect
on the $Q^2$ distributions at $Q^2 < 0.2$ GeV$^2$, and must therefore be taken 
into account.

In conclusion, NMBT provides a fully consistent and computationally viable 
scheme to calculate the electroweak nuclear response. Using the approach 
discussed in this paper may greatly contribute to decrease the systematic 
uncertainties associated with the analysis of neutrino oscillation experiments, 
 as, unlike the FG model and other many-body approaches based on effective NN 
interactions (see, e.g., Ref. \cite{nieves} and references therein), it is 
strongly constrained by NN data and involves no adjustable parameters.

\acknowledgments
This work is supported by the U. S. Department of Energy under grant
DE-FG03-87ER40347 at CSUN and by the U. S. National Science Foundation
under grant 0244899 at Caltech.
One of the authors (OB) is deeply indebted to Vijay Pandharipande and 
Ingo Sick for a number of illuminating discussions on issues relevant to the 
subject of this work. Thanks are also due to Steven Pieper for providing 
Monte Carlo configurations sampled from the oxygen ground state wave function
of Ref. \cite{16Owf}, as well as tables of the medium modified NN cross sections.



\begin{thebibliography}{99}
\bibitem{foot1}
As we will consider a target having ${\rm N}={\rm Z}={\rm A}/2$, 
the spectral functions describing proton and neutron removal 
will be assumed to be the same.
\end{thebibliography}
\end{document}